\documentclass[aps,prc,groupedaddress,showkeys,showpacs,twocolumn]{revtex4}

\usepackage{graphicx}

\begin{document}

\title{Effects of a phase transition on HBT correlations in an integrated Boltzmann+Hydrodynamics approach}

\author {Qingfeng Li$\, ^{1,2}$\footnote{liqf@fias.uni-frankfurt.de}, Jan Steinheimer$\, ^{3}$\footnote{steinheimer@th.physik.uni-frankfurt.de}, Hannah Petersen$\, ^{2,3}$\footnote{petersen@th.physik.uni-frankfurt.de}, Marcus Bleicher$\,
^{3}$\footnote{bleicher@th.physik.uni-frankfurt.de}, and Horst
St\"ocker$\, ^{2,3,4}$\footnote{H.Stoecker@gsi.de}}
\address{
1) School of Science, Huzhou Teachers College, Huzhou 313000,
China\\
2) Frankfurt Institute for Advanced Studies (FIAS), Johann Wolfgang
Goethe-Universit\"{a}t, Ruth-Moufang-Str.\ 1, D-60438 Frankfurt am
Main, Germany\\
3) Institut f\"{u}r Theoretische Physik, Johann Wolfgang
Goethe-Universit\"{a}t, Max-von-Laue-Str.\ 1, D-60438 Frankfurt am
Main, Germany\\
4) Gesellschaft f\"ur Schwerionenforschung (GSI), Planckstr. 1,
D-64291 Darmstadt, Germany
 }


\begin{abstract}
A systematic study of HBT radii of pions, produced in heavy ion
collisions  in the intermediate energy regime (SPS), from an
integrated (3+1)d Boltzmann+hydrodynamics approach is presented. The
calculations in this hybrid approach, incorporating an hydrodynamic
stage into the Ultra-relativistic Quantum Molecular Dynamics
transport model, allow for a comparison of different equations of
state retaining the same initial conditions and final freeze-out.
The results are also compared to the pure cascade transport model
calculations in the context of the available data. Furthermore, the
effect of different treatments of the hydrodynamic freeze-out
procedure on the HBT radii are investigated. It is found that the
HBT radii are essentially insensitive to the details of the
freeze-out prescription as long as the final hadronic interactions
in the cascade are taken into account. The HBT radii $R_L$ and $R_O$
and the $R_O/R_S$ ratio are sensitive to the EoS that is employed
during the hydrodynamic evolution. We conclude that the increased
lifetime in case of a phase transition to a QGP (via a Bag Model
equation of state) is not supported by the available data.
\end{abstract}

\keywords{Dynamic transport, HBT correlation, equation of state}

 \pacs{25.75.Gz,25.75.-q,24.10.Lx}

\maketitle

\section{Introduction}
One of the main purposes of the research in heavy ion collisions
(HICs) at high beam energies is to explore the existence of the
quark gluon plasma (QGP) as well as its properties. The equation of
state (EoS) of nuclear matter is one of the key points to gain
further understanding since the EoS directly provides the
relationship between the pressure and the energy at a given
net-baryon density. Phase transitions (PT), e.g., from the hadron
resonance gas phase (HG) to the color-deconfined QGP (see e.g.,
\cite{Rischke:1995mt,Spieles:1997ab,Bluhm:2007nu}), constitute
themselves in changes of the underlying EoS.

Although, on the low temperature side (and for low baryo-chemical
potentials $\mu_B$), investigations of the EoS of nuclear matter
have been pursued for many years and uncertainties have been largely
reduced, on the high temperature side, the EoS of hot and dense QCD
matter is still not precisely known. For systems created in the RHIC
energy region with high temperatures and low baryo-chemical
potential, lattice quantum chromodynamics (lQCD) (see, e.g., Ref.\
\cite{Fodor:2001pe}) calculations predict a cross-over transition
between the hadron gas and the QGP phase. The additional structures
of the phase diagram are still under heavy debate, especially
regarding the existence or non-existence of a critical endpoint
\cite{deForcrand:2006pv}.

The intermediate SPS energy regime still raises a lot of interest
because the onset of deconfinement is expected to occur at those
energies and the possibility of a critical endpoint and a
first-order phase transition is not yet excluded. Several
beam-energy dependent observables such as the particle ratios
\cite{Afanasiev:2002mx,Alt:2007fe}, the flow
\cite{Kolb:2000fha,Bleicher:2000sx,Petersen:2006vm}, the HBT
parameters \cite{Rischke:1996em,Adamova:2002ff,Li:2007yd} show a
non-monotonic behaviour around $E_{\rm lab}=30-40A~$GeV and the
interpretation remains still unclear. Therefore, future energy scan
programs at RHIC, SPS and FAIR are planned to explore the
high-$\mu_B$ region of the phase diagram in more detail.

To learn something about the hot and dense stage of the collision
from the final state particle distributions, a dynamical modeling of
the whole process is necessary. Some of the important ingredients
which have to be considered in a consistent manner are

\begin{itemize}
\item
the initial conditions and the initial nonequilibrium dynamics,
\item
the treatment of the phase transition and hadronization, as well as
the right degrees of freedom,
\item
viscosity effects in the initial partonic as well as in the hadronic
stage of the evolution,
\item
hadronic rescatterings and freeze-out dynamics.
\end{itemize}
We notice that part of these have been pointed out to be of
importance especially for the understanding of the HBT results
\cite{Pratt:2008sz,Lisa:2008gf,Broniowski:2008vp}.

Combined microscopic+macroscopic approaches are among the most
successful ideas for the modeling of the bulk properties of HICs
\cite{Andrade:2006yh,Hirano:2005xf,Nonaka:2006yn}. Recently, a
transport approach that embeds a full (3+1) dimensional ideal
relativistic one fluid evolution for the hot and dense stage of the
reaction has been developed and first results are convincing
\cite{Steinheimer:2007iy,Petersen:2008dd}. This hybrid model
inherits the advantages of the Ultra-relativistic Quantum Molecular
Dynamics (UrQMD) model for the dynamic treatment of the initial and
the final state by taking into account event-by-event fluctuations.
Furthermore, the hybrid model allows for a dynamical coupling
between hydrodynamics and transport calculation in such a way that
one can compare calculations with various EoS during the
hydrodynamic evolution and with the pure cascade calculations within
the same framework.

It is well-known that by using HBT interferometry techniques one can
get detailed information about the space-time configuration of the
system at freeze-out. We concentrate here on the two (identical)
pion interferometry and test the sensitivity of the HBT results on
different stages of the evolution. In our previous investigations on
the HBT correlation of various identical particle pairs from HICs at
AGS, SPS, and RHIC energies
\cite{Li:2006gp,Li:2006gb,Li:2007im,Li:2007yd,Li:2008ge,Li:2008bk},
we adopted the UrQMD model but further considered the mean field
potentials for both confined and ``pre-formed'' particles in the
model \cite{Li:2007yd,Li:2008ge,Li:2008bk}. We found that also
initial stage interaction may contribute to a better description of
the HBT time-related puzzle throughout the energies from AGS, SPS,
up to RHIC.

In this paper we perform a systematic investigation of the
sensitivity of HBT correlation of negatively charged pions to the
EoS by applying the newly developed hybrid approach. Similar more
focused studies were frequently discussed with microscopic transport
or hydrodynamic models before
\cite{Rischke:1996em,Soff:2000eh,Zschiesche:2001dx,Lisa:2005dd}. It
is also interesting to study if the current set of EoS employed in
the hydrodynamic phase support the conclusion about the origin of
the HBT time-related puzzle. In addition, the effects of the
hadronic rescattering and of resonance decays (dubbed as ``HR'')
after the hydrodynamic freeze-out on the HBT radii and the $R_O/R_S$
ratio deserve more investigation. We have also noticed that some
recent progresses of this topic both from an improved hydrodynamic
calculation \cite{Broniowski:2008vp} and from the
pion-optical-potential point of view \cite{Luzum:2008tc} have been
published which provides additional new insights.

The paper is arranged as follows. In the next section, the
UrQMD+hydrodynamics hybrid model is introduced briefly. The set of
different EoS that are employed in the hydrodynamic phase are
explained. Two different treatments for the transition process from
the hydrodynamic evolution to the final state hadronic cascade are
discussed. In Section 3, the analyzing program CRAB for constructing
the HBT correlator and the corresponding three dimensional (3D)
Gaussian fitting process are introduced. In Section 4, the HBT radii
$R_L$, $R_O$, and $R_S$, and the $R_O/R_S$ ratio of the negatively
charged pion source from central Pb+Pb collisions at SPS energies
are shown and discussed in the context of the experimental data.
Finally, in Section 5, a summary and an outlook are given.

\section{UrQMD+hydrodynamic model}

An integrated Boltzmann+hydrodynamics transport approach is applied
to simulate the dynamics of the heavy ion collision. To mimic
experimental conditions as realistic as possible the initial
conditions and the final hadronic freeze-out are calculated using
the UrQMD approach. Especially for an observable like HBT radii it
is important to take care of the complexity of the different effects
\cite{Pratt:2008sz,Lisa:2008gf}. The non-equilibrium dynamics, e.g.
fluctuations of the local baryon and energy density
\cite{Bleicher:1998wd}, in the very early stage of the collision and
the final state hadronic interactions are properly taken into
account on an event-by-event-basis.

UrQMD is a microscopic transport approach based on the covariant
propagation of constituent (anti-)quarks and diquarks accompanied by
mesons and baryons, as well as the corresponding anti-particles,
i.e., full baryon-antibaryon symmetry is included. It simulates
multiple interactions of ingoing and newly produced particles, the
excitation and fragmentation of color strings
\cite{NilssonAlmqvist:1986rx,Andersson:1986gw,Sjostrand:1993yb} and
the formation and decay of hadronic resonances
\cite{Bass:1998ca,Bleicher:1999xi}. In principle it is also possible
to incorporate mean field interactions in the transport calculation,
In the present calculation they are neglected in order to test the
hydro-phase and in the following we will refer to the pure cascade
calculation as UrQMD-2.3. Studies on the thermodynamic properties of
UrQMD can be found in
\cite{Bass:1997xw,Bravina:1998it,Bravina:2008ra}.

The coupling between the UrQMD initial state and the hydrodynamical
evolution proceeds when the two Lorentz-contracted nuclei have
passed through each other, $t_{\rm start} = {2R}/{\sqrt{\gamma^2
-1}}$ \cite{Steinheimer:2007iy}. After that, a full (3+1)
dimensional ideal hydrodynamic evolution is performed using the
SHASTA algorithm \cite{Rischke:1995ir,Rischke:1995mt}. The
hydrodynamic evolution is stopped, if the energy density
$\varepsilon$ of all cells drops below five (default value) times
the ground state energy density $\varepsilon_0$ (i.e. $\sim 730 {\rm
MeV/fm}^3$). This criterion corresponds to a T-$\mu_B$-configuration
where the phase transition is expected - approximately $T=170$ MeV
at $\mu_B=0$. The hydrodynamic fields are then mapped to particle
degrees of freedom via the Cooper-Frye equation on an isochronous
(in the computational frame) hypersurface. The particle vector
information is then transferred back to the UrQMD model, where
rescatterings and final decays are calculated using the hadronic
cascade. We will further refer to this kind of freeze-out procedure
as the isochronous freeze-out (IF). This procedure is explained in
detail in \cite{Petersen:2008dd}.

In this paper we introduce another freeze-out procedure to account
for the large time dilatation that occurs for fluid elements at
large rapidities. Faster fluid elements need a longer time to cool
down to the same temperatures than the cells at midrapidity since
the hydrodynamic calculation is performed in the center-of-mass
frame of the collision. At higher energies the isochronous
hypersurface increasingly differs from an iso-$\tau$ hypersurface
($\tau$ is the proper time). To mimic an iso-$\tau$ hypersurface we
therefore freeze out transverse slices, of thickness $\Delta z = 0.2
$fm, whenever all cells of that slice fulfill our freeze-out
criterion. For each slice we apply the isochronous procedure
described above separately. By doing this we obtain a rapidity
independent freeze-out temperature even for the highest beam
energies. For lower energies ($E_{lab}\lesssim 80A$ GeV) the two
procedures yield very similar results for the temperature
distributions. The hydrodynamic fields are then again mapped to
particle degrees of freedom via the Cooper-Frye equation on this new
hypersurface. In the following we will refer to this procedure as
``gradual freeze-out''(GF). A more detailed description of the
hybrid model including parameter tests and results for
multiplicities and spectra can be found in \cite{Petersen:2008dd}.

Serving as an input for the hydrodynamical calculation the EoS
strongly influences the dynamics of an expanding system. In this
work we use three different EoS to investigate their effect on the
extracted HBT radii. The first EoS, named the hadron gas (HG),
describes a non-interacting gas of free hadrons
\cite{Zschiesche:2002zr}. Included here are all reliably known
hadrons with masses up to $2$ GeV, which is equivalent to the active
degrees of freedom of the UrQMD model (note that this EoS does not
contain any form of phase transition). This purely hadronic
calculation serves as a baseline calculation to explore the effects
of the change in the underlying dynamics - pure transport vs.
hydrodynamic calculation. The second EoS, named the Bag Model EoS
(BM), follows from coupling a bag model of massless quarks and
gluons to a Walecka type of hadron gas including only SU(2) flavours
(for details the reader is referred to \cite{Rischke:1995mt}). This
EoS exhibits a strong first-order phase transition (with large
latent heat) for all baryonic chemical potentials $\mu_B$. The third
EoS, named the chiral+HG (CH), follows from a chiral hadronic
$SU(3)$ Lagrangian and incorporates the complete set of baryons from
the lowest flavour-$SU(3)$ octet, as well as the entire multiplets
of scalar, pseudo-scalar, vector and axial-vector mesons
\cite{Papazoglou:1998vr}. Additional baryonic degrees of freedom are
included to produce a first-order phase transition in certain
regimes of the $T$-$\mu_q$ plane, depending on the couplings
\cite{Theis:1984qc,Zschiesche:2001dx,Zschiesche:2004si}. Using this
EoS, a phase structure including a first-order phase transition and
a critical endpoint at finite $\mu_B$ is obtained
\cite{Zschiesche:2006rf}. This EoS has already been successfully
applied to a hydrodynamic calculation \cite{Steinheimer:2007iy}.

\begin{figure}
\includegraphics[angle=0,width=0.48\textwidth]{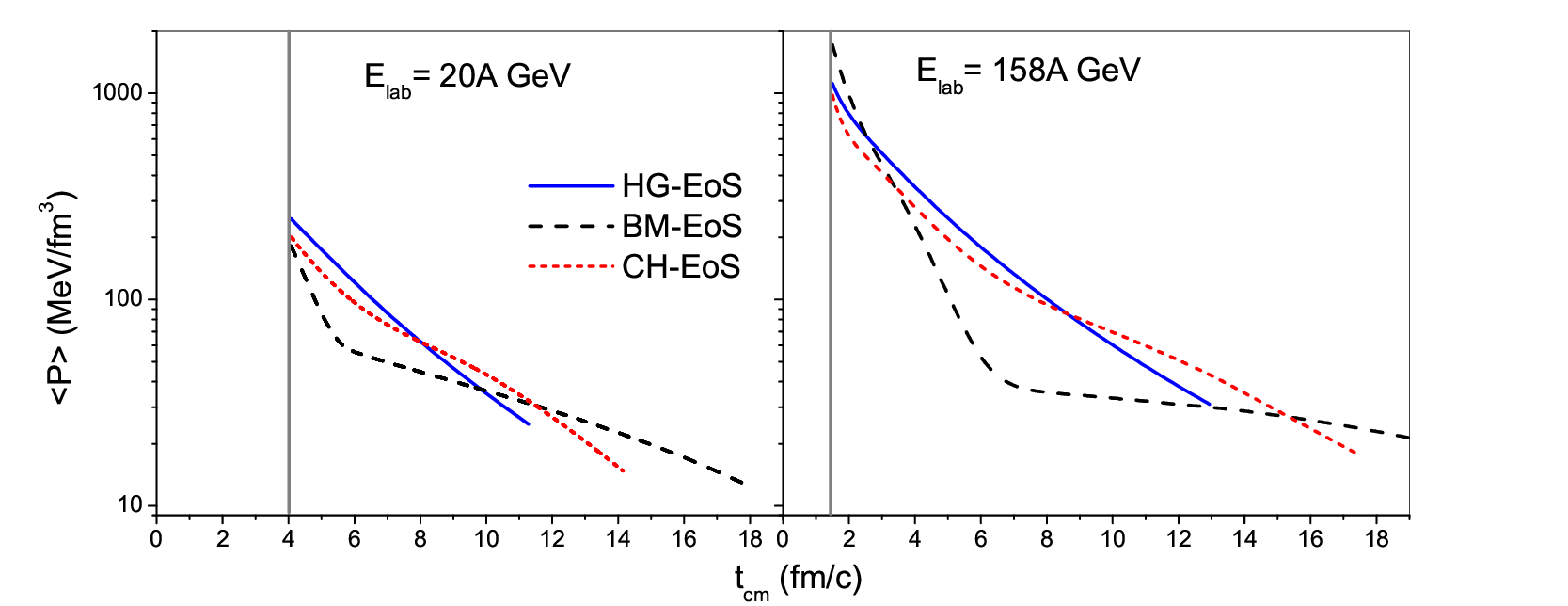}
\caption{Time (in the center of mass frame) evolution of the average
pressure for all three EoS and a central Pb+Pb collision at $E_{\rm
lab}=20A$ GeV (left plot) and $E_{\rm lab}=158A$ GeV (right plot).
The vertical line in each plot indicates the starting time of the
hydro evolution.} \label{press_t}
\end{figure}

To visualize the differences of these EoS, Fig.\ \ref{press_t} shows
the average pressure of the expanding system, from central Pb+Pb
collisions at $E_{\rm lab} = 20A$ GeV (left plot) and $158A$ GeV
(right plot), as a function of time (in the center of mass frame).
The vertical line in each plot indicates the starting time of the
hydro evolution. The mean value of the pressure has been obtained by
weighting the pressure, $P_{i}$, in every cell $i$ by its energy
density, $\varepsilon_{i}$, and integrating over the hydrodynamic
grid

\begin{equation}
<P> =\frac{\sum_{i}{ P_{i}\cdot\varepsilon_{i} }
}{\sum_{i}{\varepsilon_{i}}}.
\end{equation}
All curves in Fig.\ \ref{press_t} are plotted until the point in
time when the isochronous freeze-out criterion is fulfilled.
Compared to the HG the BM-EoS leads to a delayed freeze-out time
(i.e. a much longer expansion). While in the first few $fm/c$ of the
evolution, the system obeying the BM-EoS expands most violently (due
to the high pressure gradient in the QGP phase), once the system
enters the mixed phase, its expansion is slowed down considerably.
This can be observed as the ``kink'' in Fig.\ \ref{press_t}. At the
higher beam energy, $E_{lab} = 158A$ GeV, this softening of the EoS
is even more pronounced. Since the HG-EoS does not contain any phase
transition, no softening can be observed, resulting in the shortest
expansion time. The chiral CH-EoS lies in between both extreme
cases. Although a small kink can be observed, it is not as
pronounced as in the BM-EoS. The effect of changes in the EoS on HBT
results has been studied before \cite{Zschiesche:2001dx}, but the
great advantages of our approach are the full (3+1) dimensions and
the same initial conditions and freeze-out for all three cases and
all beam energies without adjusting additional parameters.

\section{CRAB analyzing program and the fitting process}

To calculate the two-particle correlator, the CRAB program is
adopted \cite{Pratthome}, which is based on the formula:
\begin{equation}
C({\bf k},{\bf q}) = \frac {\int d^4x_i d^4x_j g(x_i,p_i) g(x_j,
p_j) |\phi({\bf r}^\prime, {\bf q}^\prime)|^2} {{\int d^4x_i
g(x_i,p_i)} {\int d^4x_j g(x_j,p_j)}}. \label{cpq}
\end{equation}
Here $g(x_i,p_i)$ is an effective probability for emitting a
particle $i$ with 4-momentum $p_i=({E_i, \bf p}_i)$ from the
space-time point $x_i = (t_i, {\bf r}_i)$. $\phi({\bf r}^\prime,
{\bf q}^\prime)$ is the relative wave function with ${\bf r}^\prime$
being the relative position in the pair's rest frame. ${\bf q}={\bf
p}_i-{\bf p}_j$ and $\textbf{k}=(\textbf{p}_{i}+\textbf{p}_{j})/2$
are the relative momentum and the average momentum of the two
particles $i$ and $j$.

In this work, we select central ($<7.2\%$ of the total cross section
$\sigma_T$) Pb+Pb collisions at SPS energies: $E_b=20A$, $30A$,
$40A$, $80A$ and $158A$ GeV, with a pair rapidity cut
$|Y_{\pi\pi}|<0.5$ ($Y_{\pi\pi}={\rm log}((E_1+E_2+p_{\parallel
1}+p_{\parallel 2})/(E_1+E_2-p_{\parallel 1}-p_{\parallel 2}))/2$ is
the pair rapidity with pion energies $E_1$ and $E_2$ and
longitudinal momenta $p_{\parallel 1}$ and $p_{\parallel 2}$ in the
center of mass system). For each EoS about $2500$ events are
calculated. All particles with their phase space coordinates at
freeze-out are then given into the CRAB analyzing program. Only the
negatively charged pions are considered during the analyzing process
(for each analysis, one hundred million pion pairs are considered).
For the cascade calculations, we take the results from our previous
publications as reference \cite{Li:2006gb,Li:2007yd}. We found that
the residual Coulomb effect after the hadron freeze-out on the HBT
radii of the pion source is small \cite{Li:2008ge}, therefore we
omit it in the present analysis. Finally, we choose the longitudinal
comoving system (LCMS) frame of the pair (also called the
``Out-Side-Long'' system, in which the longitudinal component of the
pair velocity vanishes), which is frequently adopted in recent
years, and fit the correlator by a 3D Gaussian distribution

\begin{eqnarray}
&&C(q_O,q_S,q_L)= K[1+\lambda \nonumber \\
&&\times {\rm
exp}(-R_L^2q_L^2-R_O^2q_O^2-R_S^2q_S^2-2R_{OL}^2q_Oq_L)].
\label{fit1}
\end{eqnarray}
Here $K$ is the overall normalization factor, the $q_x$ and $R_x$
are the components of the pair relative momentum and homogeneity
length (HBT radius) in the $x$ direction, respectively. The
$\lambda$ parameter is called as the incoherence factor or, more
correctly, the intercept parameter and lies for Bose-Einstein
statistics between 0 and 1 for two-boson correlations in realistic
HICs. Because the parameter $\lambda$ might be influenced by many
additional factors, such as contamination, long-lived resonances, or
the details of the residual Coulomb modification, we regard it as a
free parameter and do not show it in this letter. However, In
\cite{Li:2008ge} we have found that the calculated $\lambda$ factor
with UrQMD can be comparable with experimental data at RHIC energies
(although somewhat larger than data). At SPS energies, $\lambda_{\rm
UrQMD}\approx 0.8-0.9$, is also compatible with experimental data
which $\lambda_{\rm data}\approx 0.6-0.8$. The $R_{OL}^{2}$
represents the cross-term and plays a role at large rapidity. To fit
the correlator with Eq.~(\ref{fit1}), we use ROOT
\cite{roothomepage} software and minimize $\chi^2$.

\section{HBT results}

\begin{figure}
\includegraphics[angle=0,width=0.48\textwidth]{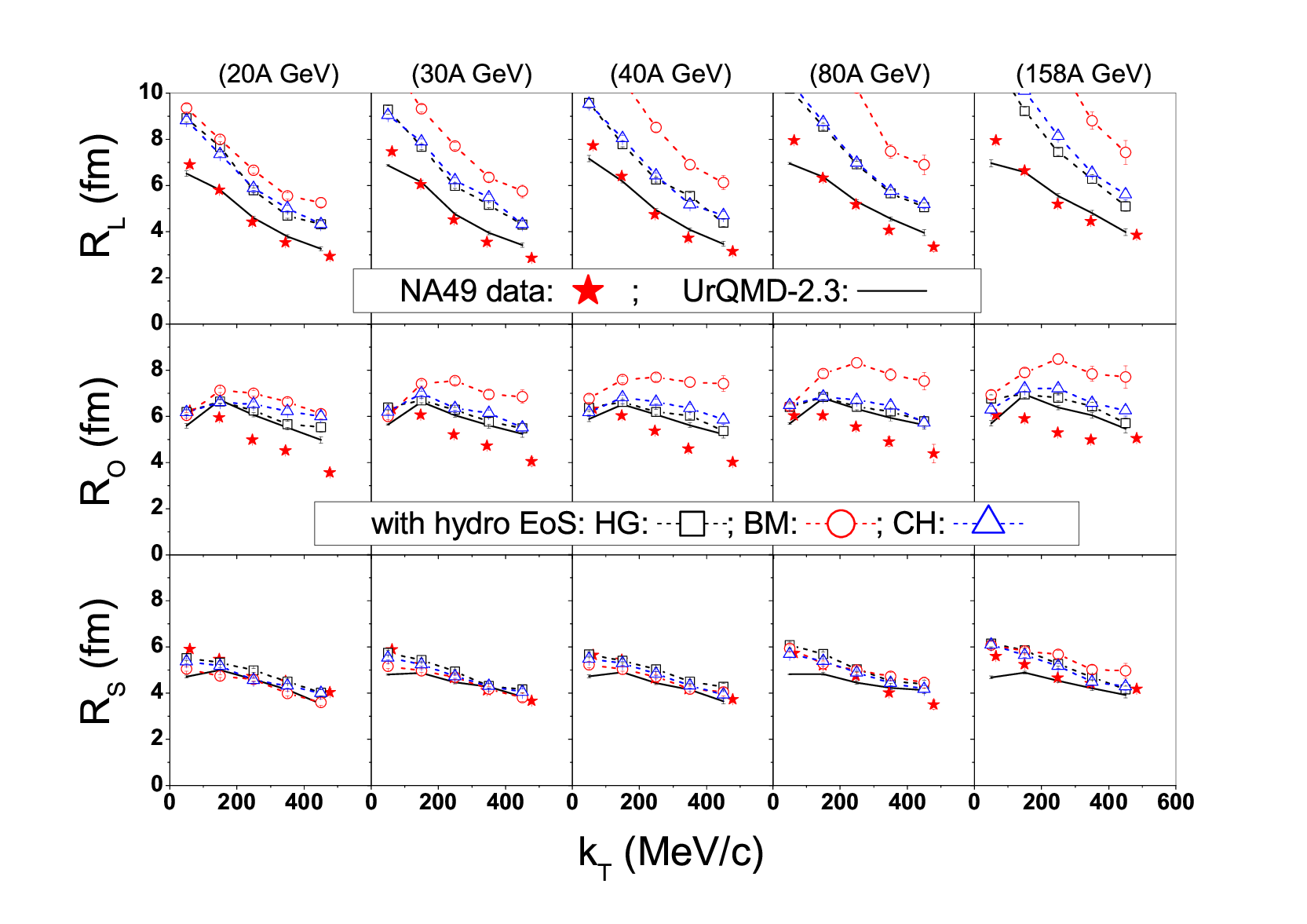}
\caption{Transverse momentum $k_T$ dependence of the HBT radii
$R_L$, $R_O$, and $R_S$ (at midrapidity) of $\pi^-$ source from
central HICs at SPS energies ($E_{\rm lab}=20A$, $30A$, $40A$,
$80A$, and $158A$ GeV). The NA49 data are indicated by solid stars
\cite{Alt:2007uj}. The pure cascade calculation is depicted by lines
while the hybrid model calculations with different EoS (HG, BM and
CH) are depicted by dashed lines with open symbols. }
\label{HBT-fig1}
\end{figure}

\begin{figure}
\includegraphics[angle=0,width=0.48\textwidth]{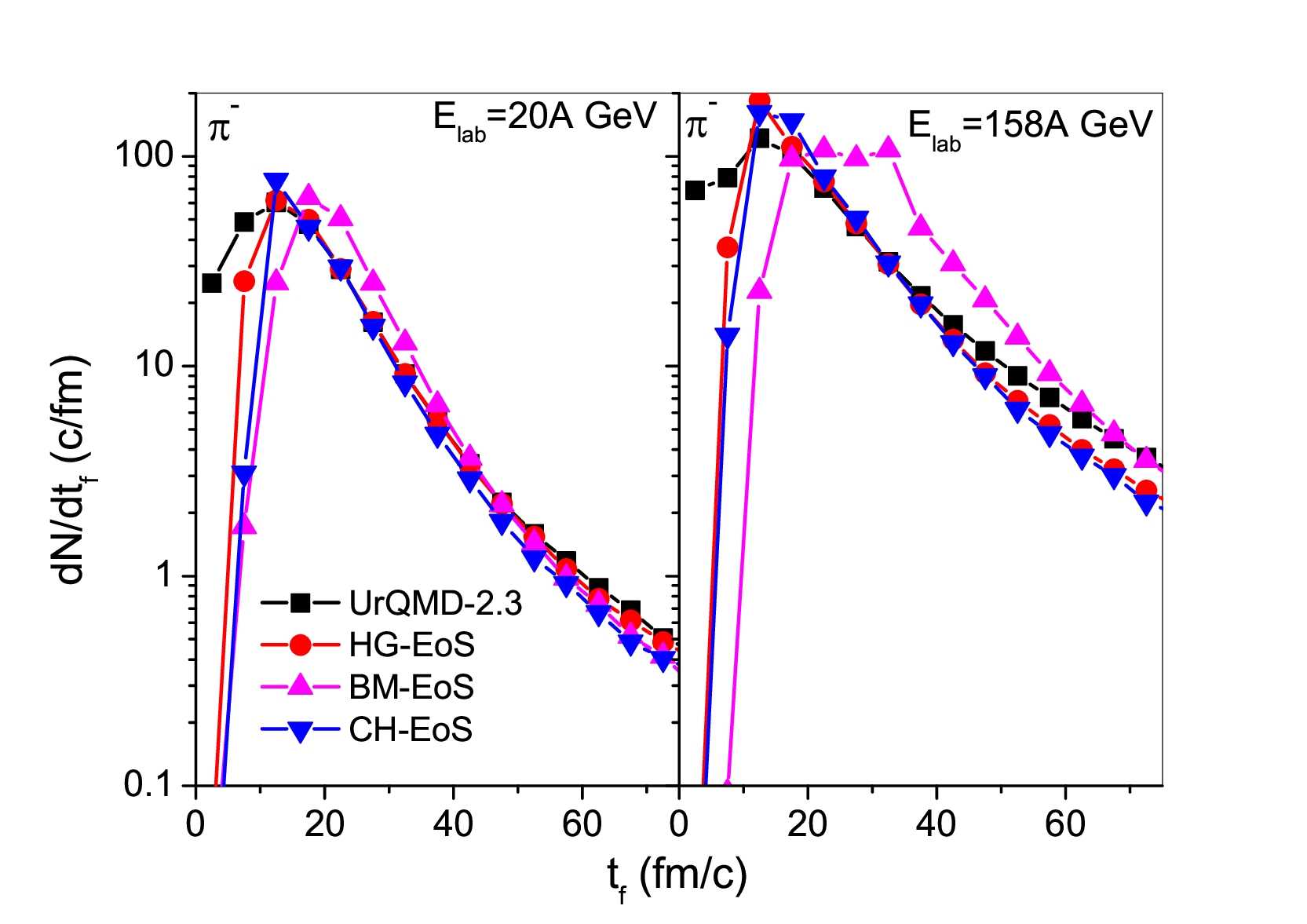}
\caption{Freeze-out time dependence of the $\pi^-$ emission in
central Pb+Pb at $E_{\rm lab}=20A$ GeV (left plot) and $158A$ GeV
(right). Calculations with the UrQMD cascade are compared with the
hybrid model calculations with the EoS of HG, BM, and CH. }
\label{HBT-fig2}
\end{figure}

Fig.\ \ref{HBT-fig1} shows the transverse momentum $k_T$
($\textbf{k}_T=(\textbf{p}_{1T}+\textbf{p}_{2T})/2$) dependence of
the HBT radii $R_L$, $R_O$, and $R_S$ (at midrapidity) of $\pi^-$
source from central Pb+Pb collisions at SPS energies. The data
(solid stars) are from the NA49 Collaboration \cite{Alt:2007uj}. The
pure cascade calculation is depicted by lines while the hybrid model
calculations with different EoS (HG, BM and CH) are depicted by
dashed lines with open symbols. As was shown before, the cascade
calculation gives a fairly good result of the $k_T$ dependence of
$R_L$ and $R_S$ values except at quite small $k_T$, while for $R_O$,
it is slightly larger than data at large $k_T$. In contrast, the
hybrid model calculations show large HBT (for all employed EoS, but
to a varying degree) in all directions, especially in the
longitudinal direction. The HG and CH are moderately increased and
lead to very similar results for all three directions. The large
latent heat in the bag model leads to a further strong increase in
the longitudinal direction and in the transverse direction at large
$k_T$. This increase in the BM mode becomes more pronounced at
higher beam energies. At first glance, this result might be
surprising because at least in the transverse direction one would
expect a faster expansion including a hydrodynamic evolution. On the
other hand, one knows that the system spends a longer time without
emitting any particles in the hybrid model calculation.

Fig.\ \ref{HBT-fig2} exhibits the freeze-out time dependence of the
$\pi^-$ emission in central Pb+Pb at $E_{\rm lab}=20A$ GeV (left
plot) and $E_{\rm lab}=158A$ GeV (right plot). It is clearly seen
that there are almost no pions emitted before $\sim 10$fm$/c$ in the
hybrid model calculations. This is easy to understand because even
in the gradual hydro-freeze-out which is applied here, it takes a
while until the first slices have cooled down and are frozen out
from the hydrodynamic evolution. There is no particle emission from
earlier times in contrast to the pure cascade calculation. For the
BM-EoS, this effect is present even for a longer time since the
expansion lasts longer \footnote{This might change, if a continuous
emission approach is used for the hydrodynamic model e.g. suggested
in \cite{Grassi:1994nf,Knoll:2008sc}.}. The on-the-average longer
freeze-out time leads to an apparently larger size of the pion
source. Furthermore, it is clear (and expected) that the EoS with
larger latent heat (such as in the BM mode) leads to a longer
emission duration of the particles (as seen in Fig.\ \ref{HBT-fig2}
when $t_f\gtrsim 15$fm$/c$) so that it produces larger HBT radii.
This behaviour is clearly seen in the more time-dependent directions
$R_L$ and $R_O$. This behaviour might be improved by allowing
particles also to freeze out and fly into the detector at all times
of the collision. Especially, if there are fast pions produced
during the first hard collisions in UrQMD at the edge of the system
they should be able to fly into the detector without being forced
into the hydrodynamic evolution.

\begin{figure}
\includegraphics[angle=0,width=0.48\textwidth]{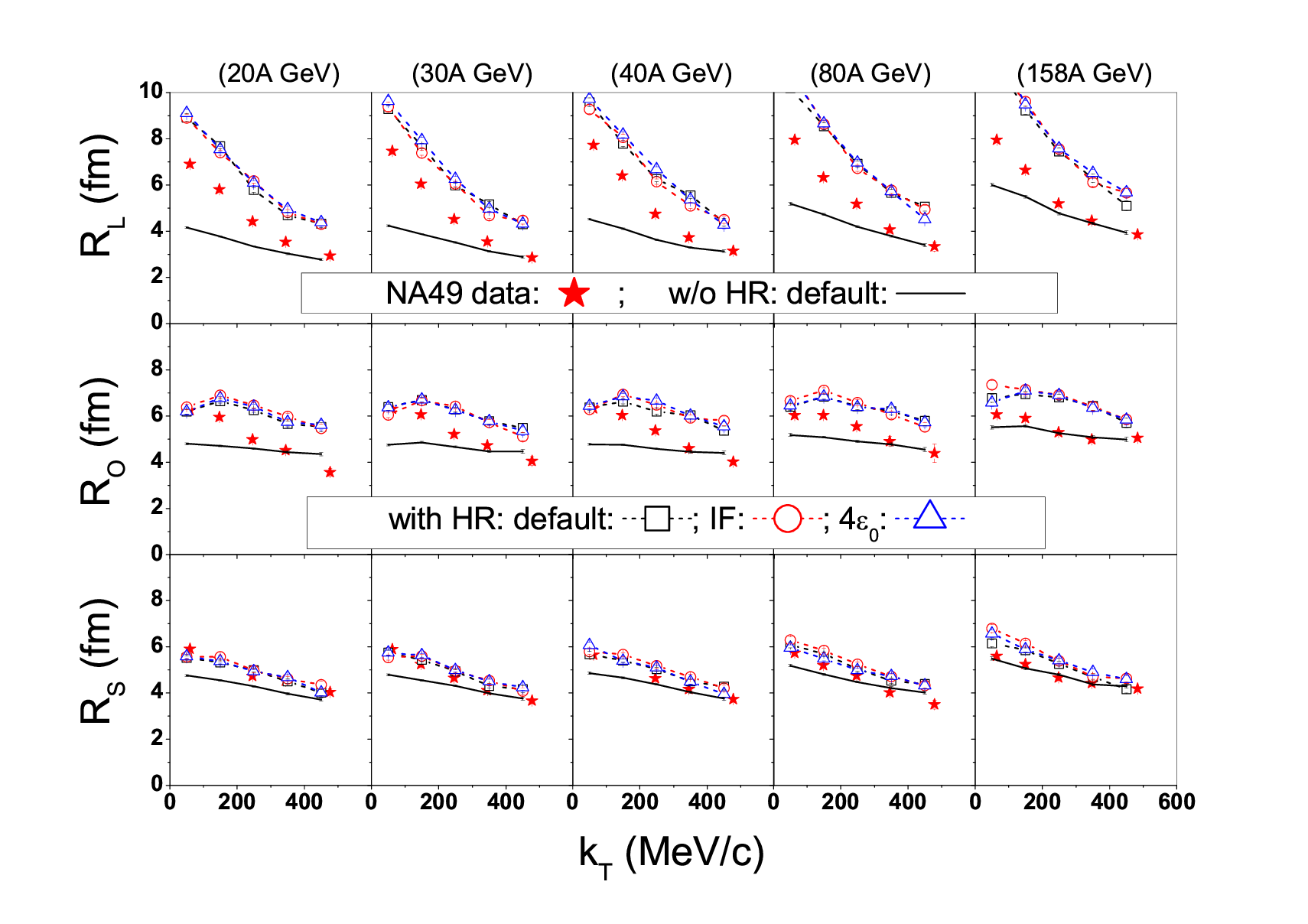}
\caption{$k_T$ dependence of the HBT radii $R_L$, $R_O$, and $R_S$
(at midrapidity) for central HICs at SPS energies ($E_{\rm
lab}=20A$, $30A$, $40A$, $80A$, and $158A$ GeV). The NA49 data are
indicated by solid stars \cite{Alt:2007uj}. The HG-EoS is employed
in all calculations but under different freeze-out conditions: 1)
without HR, calculations with default hydro-freeze-out criteria (GF,
$5\cdot\varepsilon_0$) are depicted by lines. 2) full hybrid model
calculations with two different cuts of the energy density (default
and $4\cdot \varepsilon_0$) for the GF and with the default cut of
the energy density for the IF are depicted by dashed lines with
symbols.} \label{HBT-fig3}
\end{figure}

Fig.\ \ref{HBT-fig3} illustrates the $k_T$ dependence of the HBT
radii under various freeze-out conditions, which may be divided into
two parts: 1) without HR and 2) with HR after the hydrodynamic
phase. Here, without HR (lines) means that the evolution is stopped
immediately after the Cooper-Frye freeze-out from the hydrodynamic
phase (GF and $5\cdot\varepsilon_0$ are adopted as default
hydro-freeze-out criteria), with instantaneous resonance decays. The
observed size of the pion source is small at this hydrodynamic
freeze-out. In previous investigations \cite{Petersen:2008dd} it has
been found that binary baryon-meson and meson-meson collisions still
frequently happen after the hydrodynamic freeze-out. In baryon-meson
reactions, the most abundant interactions are the excitation and the
decay of the $\Delta$ resonance (i.e. $\Delta \rightleftharpoons \pi
N$), while in meson-meson collisions, the $\rho \rightleftharpoons
\pi\pi$ process is dominant. A large number of these final hadron
interactions in which pions are involved contribute significantly to
the final HBT radii of pions in this model.

Let us therefore explore if the finally observed HBT radii do depend
on the transition criterion from hydrodynamics to the transport
model. The full hybrid model calculations (dashed lines with open
symbols) are shown with two different cuts of the energy density
($5\cdot \varepsilon_0$ as default and $4\cdot \varepsilon_0$) for
the GF and with the energy density cut $5\cdot \varepsilon_0$ for
the IF. It is found that the final state hadronic interactions are
sufficient that the effects of different treatments of the
hydrodynamic freeze-out on the final HBT radii are almost totally
washed out in all directions and at all investigated energies.

\begin{figure}
\includegraphics[angle=0,width=0.48\textwidth]{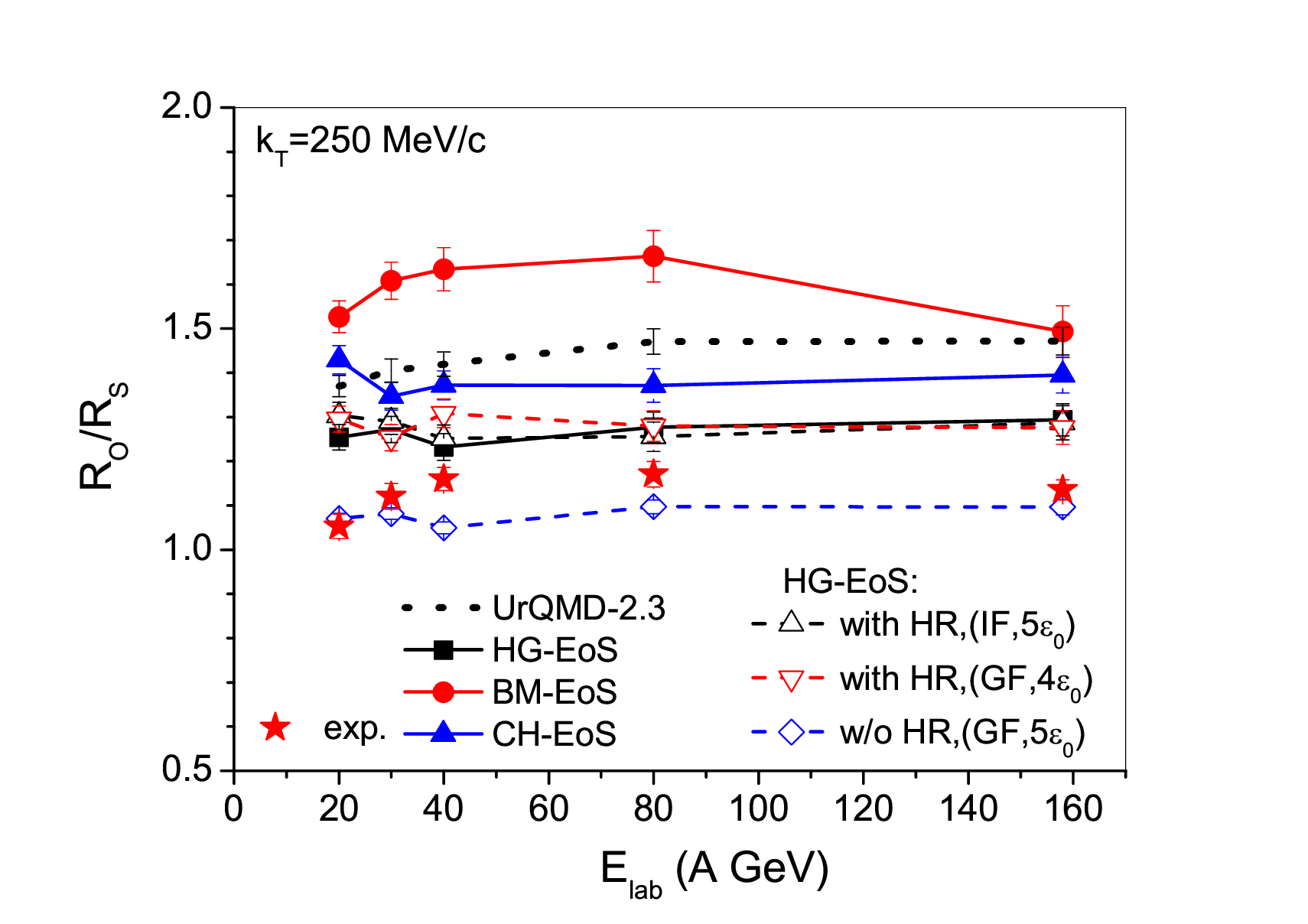}
\caption{Excitation function of the $R_O/R_S$ ratio at
$k_T=250$MeV$/c$. The NA49 data are indicated by solid stars
\cite{Alt:2007uj}. UrQMD cascade calculation is shown by dotted
line. Hybrid model calculations with EoS of HG, BM, and CH and with
HR are shown by lines with solid symbols (the default
hydro-freeze-out criteria (GF, $5\cdot\varepsilon_0$) are used). In
the HG-EoS mode, various criteria of freeze-out, (IF, $5\cdot
\varepsilon_0$) and (GF, $4\cdot \varepsilon_0$) with HR, and the
default (GF, $5\cdot\varepsilon_0$) without HR are shown by dashed
lines with open symbols.} \label{HBT-fig4}
\end{figure}

Let us finally explore the dependence of the $R_O/R_S$ ratio on the
different EoS and freeze-out prescriptions. This ratio was expected
to be sensitive to the duration time of the homogeneity region. In
Fig. \ref{HBT-fig4} the excitation function of the $R_O/R_S$ ratio
with the different EoS (lines with solid symbols) and freeze-out
prescriptions (dashed lines with open symbols) are shown. The $k_T$
bin $200-300$MeV$/c$ is chosen. The result for the pure cascade
calculation is also shown as a baseline (dotted line). It is seen
clearly that the $R_O/R_S$ ratio is sensitive to the EoS, but not to
the various hydrodynamic freeze-out prescriptions when including HR
(shown as open triangles and open inverted triangles) as it has
already been implied from the results of the HBT radii shown in
Figs.\ \ref{HBT-fig1} and \ref{HBT-fig3}. With increasing latent
heat which corresponds to the softness of EoS implied from Fig.\
\ref{press_t}, the $R_O/R_S$ ratio is increased. The ``excessively''
large latent heat in BM-EoS results in a long duration time of the
pion source and hence a large $R_O/R_S$ ratio. Although the overall
height is largely overpredicted by the BM-EoS, the qualitative
behaviour of the data (with a maximal lifetime at beam energies
around $40-100A$ GeV) is well reproduced. In addition, the ``peak
structure'' is less pronounced than in previous predictions
\cite{Rischke:1996em}, due to the different initial state and seems
to provide a more reasonable estimate of the magnitude of the
lifetime enhancement. The chiral EoS CH exhibits a lower $R_O/R_S$
ratio because the first-order phase transition is less pronounced.
The calculation with HG mode (line with solid squares) leads to the
smallest $R_O/R_S$ ratio due to the most stiffest EoS among the
three ones. The result of the cascade calculation lies in between
the CH and the BM modes, which implies a relatively soft EoS. It can
be understood since in the pure UrQMD model the new particle
production is treated either as a resonance decay or a fragmentation
of the string, which introduces a finite lifetime and hence leads to
a softer EoS. After considering the mean field potentials for both
confined and ``pre-formed'' particles \cite{Li:2007yd,Li:2008ge},
which gives a strong repulsion at the early stage, the $R_O/R_S$
ratio was seen to decrease in line with results obtained here.

For the full hybrid model calculation the different freeze-out
prescriptions do not affect the final results (when comparing the
results by dashed lines with open triangles with that by the line
with solid squares). The calculation with HG-EoS but without HR
(dashed line with open diamonds) seems to provide better description
of the data, but, seems clearly unphysical to the authors as a
solution to the duration time problem.

\section{Summary and outlook}

In summary, the HBT correlation of the negatively charged pion
source created in central Pb+Pb collisions at SPS energies was
investigated with a hybrid model that incorporates a (3+1)d
hydrodynamic evolution in the UrQMD transport approach. We explored
different settings, one where the EoS was varied without changing
the initial conditions and the freeze-out prescription and another
where the EoS was fixed and the treatment of the freeze-out was
changed. We presented a systematic investigation of these effects on
the HBT radii. It was found that the latent heat influences the
emission of particles visibly and hence the HBT radii of the pion
source. While the final rescatterings result in an independence of
the calculated HBT-parameters from the transition criterion, they do
not improve the quantitative agreement with the experimental data.
The details of the hydro-freeze-out prescription do not affect the
HBT radii as well as the $R_O/R_S$ ratio as long as the HR in the
subsequent hadronic transport model were taken into account.
Overall, the HBT data seem to favor a stiff EoS
\cite{Bekele:2007ee}, but one should also keep in mind that
viscosity effects are neglected during the hydrodynamic stage and
that the particle emission from the early stages should be handled
more carefully.

In the future, bulk and shear viscosity will be further considered
for the hydrodynamic phase, and the non-equilibrated particle
emission should be treated more precisely.

\section*{Acknowledgments}
We thank S. Pratt for providing the CRAB program and thank D.H.
Rischke for providing the hydrodynamics code. We acknowledge support
by the Frankfurt Center for Scientific Computing (CSC). This work
was supported by the Hessian LOEWE initiative through the Helmholtz
International Center for FAIR (HIC for FAIR) and GSI and BMBF. H.P.
acknowledges financial support from the Deutsche Telekom Stiftung
and support from the Helmholtz Research School on Quark Matter
Studies.


\end{document}